Elizabeth Locci (elizabeth.locci@cern.ch)    02/06/2015

# ASAHEL
## (A Simple Apparatus for High Energy LEP)

# Proposal[1]

E. Locci (CEA/DSM/IRFU/SPP, Gif-sur-Yvette, France)

## Abstract

The TLEP Design Study Working Group published "First Look at the TLEP Physics Case" in December 2013. TLEP, a 90-400 GeV high-luminosity, high precision, $e^+e^-$ machine, is now part of the Future Circular Collider (FCC) design study, as a possible first step (named FCC-ee) towards a high-energy proton-proton collider (named FCC-hh).

The above paper presents an initial assessment of some of the relevant features of the FCC-ee potential, to serve as a starting point for the more extensive design study that is now carried out.

FCC-ee will provide the opportunity to make the most sensitive tests of the Standard Model of electroweak interactions. The first requirement of the detector must therefore be to ensure it has the capability to make these precise tests. The detector must have excellent vertexing and tracking performances and a highly granular, homogeneous calorimetric system covering as great a solid angle as possible.

Following the ALEPH philosophy of using as few detection techniques as possible, we propose to evaluate the modifications that would be needed for an "ALEPH-like" detector to fulfil the requirements of FCC-ee. We will investigate the use of Micromegas detectors instead of limited-streamer tubes and will adjust the size of the detector and its granularity.

---

[1] This document is not meant for publication in the present state, but should be considered as a basis for future work to explore the adequacy of an « ALEPH-like » detector with the physics benchmarks of FCC-ee.

# Table of contents

# 1. Introduction

The proposed detector should be designed for an experiment using the conceptual $e^+e^-$ collider (FCC-ee) capable of very high luminosities ($> 10^{35}$ cm$^2$s$^{-1}$) in a wide centre-of-mass ($E_{CM}$) spectrum from 90 to 350 GeV.

The physics case of FCC-ee is described in reference [1]. The detailed physics program will depend on the results of the LHC run at 13-14 TeV. A strategy to look beyond the LHC findings would include the measurement of Z, W, top and Higgs properties with sufficient accuracy to provide sensitivity to new physics at a much higher energy scale, as well as direct search for new physics.

The experimental conditions of FCC-ee will be similar to those of LEP, with the additional bonus of very stable beam conditions brought by the continuous top-off injection. The events will be as "clean" as at LEP, with no pile-up interactions and negligible beam background.
However a number of differences in the machine parameters will have some impact on the design of the detectors:

- A smaller value of the vertical β function at the interaction point (IP) increases the beam divergence at the IP and may have a sizeable effect on the acceptance of low angle detectors used for the luminosity measurement.

- The strong final focus quadrupoles will generate large amounts of synchrotron radiation against which appropriate shielding must be provided.

- More beamstrahlung may lead to larger background of electromagnetic radiation at the IP, nevertheless several orders of magnitude smaller than at a linear collider.

- The run at the Z pole with a luminosity of $5 \times 10^{35}$ cm$^{-2}$s$^{-1}$ at each interaction point, corresponds to a trigger rate of 15 kHz for Z decays in the central detector and 60 kHz for Bhabha scattering in the luminometer. The detectors will have to sustain this high trigger rate.

- The repetition rate will reach 20 MHz at the Z pole (LEP operated at 44 kHz). The rate capability of the detectors will be required to match this high repetition rate.

The essential elements are
- a vertex detector, having performances similar to those required for a linear collider detector, with lifetime-based c-tagging capabilities,
- a very good tracking capability
- a highly granular, homogeneous, calorimetric system covering as great a solid angle as possible, for particle-flow purposes,
- a precision device for luminosity measurement with Bhabha scattering.

The document is split into three parts. In the first part the main characteristics of the LEP, LHC, and ILC detectors[2] will be investigated. In the second part the performances of the LEP and LHC experiments in the measurements of some selected observables are discussed. In the third part, conclusions from part one and two are drawn to define the proposed detector for FCC-ee.

## 2. Overview of the LEP, LHC, and ILC Detectors

For the general concept of a detector for FCC-ee, the inspiration should come from our past experience with the LEP detectors, mitigated with the recent developments for LHC, ILC.

Most numbers are taken from Letters of Intent or Technical Design Reports (ALEPH [2], DELPHI [3], L3 [4], OPAL [5], ATLAS [6], CMS [7], ILC [8]), but some were available in various publications only.

Based on the comparison of the characteristics and performances of the LEP detectors, the following observations can be made:

- the design philosophy of the ALEPH detector was driven by the early decision to use as few different detection techniques as possible.
- DELPHI was designed to provide high granularity over $4\pi$ solid angle, allowing effective particle identification, and multiplied the number of detection techniques.
- L3 concentrated its efforts on limited goals of measuring photons, electrons, and muons with high resolution. This choice was relevant for the precise measurement of leptonic decays of the Z boson, but perhaps not the best for the measurement of jets and the potential discovery of the Standard Model Higgs that decays mainly into bb quark pairs (74% for a mass of 115 GeV) [9]. An excellent measurement of photons might have been a clever choice for the discovery of a phermiophobic Higgs, mainly decaying to a pair of photons [10].
- OPAL was motivated by the exploration of an unknown energy region of $e^+e^-$ collisions with an optimum detector, using at the time well-proven technologies.

Since the design of LEP detectors in the 1990's, the considered detection techniques have improved and new detection techniques have developed, and an ALEPH-like design has to evolve with them. As a guideline for this some lessons can be taken from the design and operation of the LHC detectors and from the studies for ILC facilities.

In the following sub-sections, the different sub-detectors will be examined.

---

[2] LEP experiments have been chosen as the last detectors having recorded data at lepton colliders. Tevatron experiments have been discarded, as LHC experiments are the last generation of detectors operating at hadron colliders. At last ILC experiments are considered as they benefit from the latest advances of technology, and they have several common points with FCC-ee experiments, although some differences may arise from different machines.

### 2.1 The Magnet

LEP, LHC, and ILC experiments have chosen solenoids, and the ATLAS experiment has completed the magnet system with a toroid.

In a solenoid producing a cylindrically symmetrical field, which axis coincides with that of the colliding beams, sagitta S of the trajectory of a charged particle of momentum p, emanating at azimuthal angle $\theta = 90°$ from the interaction is:

$$S = 0.0375 \, B \, (R_{outer} - R_{inner})^2 / p$$

leading to an expected momentum resolution:

$$\Delta p / p = p \, \Delta s / [0.0375 \, B \, (R_{outer} - R_{inner})^2]$$

where $R_{outer}$, $R_{inner}$, $\Delta s$ are respectively the inner and outer radius of the solenoid, and the expected precision on the sagitta as measured by the tracker.
At $\theta < 90°$, the potential resolution improves by a factor of about $\sin \theta$.

If the total length of the solenoid is L, the fraction of the solid angle covered (Figure 1) is:

$$d\Omega / 4\pi = \cos \theta = \alpha / \sqrt{(4 + \alpha^2)}, \text{ where } \alpha = L / R \text{ is the detector aspect ratio.}$$

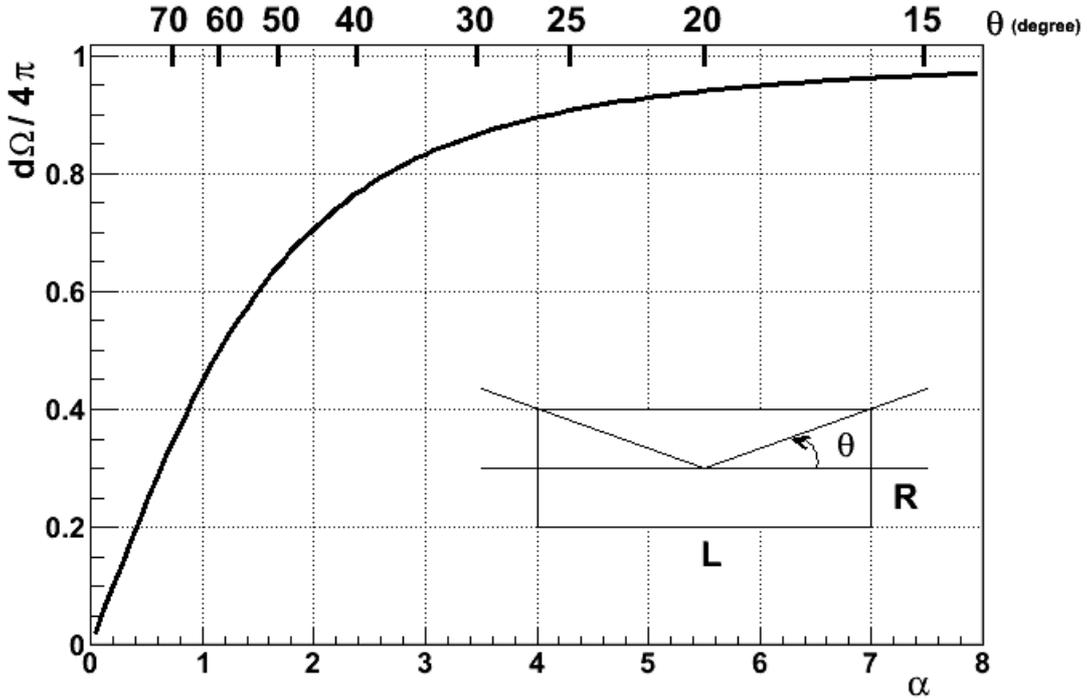

Figure 1: solid angle subtented by a cylinder of length L and radius R

The choice of the length and radius of the coil depends on the options retained for the calorimetry: only ECAL inside the coil, or both ECAL and HCAL, compactness of the calorimeters inside the coil. Both parameters have to be optimized together with the field, as the stored energy U roughly scales with $R^2 \times L \times B^2$, and is generally considered as limited to about 2.5 GJ, mainly for optimal balance between cost and required performances [11], although 50 to 100 GJ are feasible if actually needed.
Table 1 displays the choices of the LEP, LHC, ILC experiments in terms of magnetic field, coil radius and length, and resulting detector aspect ratio, solid angle, and stored energy U.

|   | **ALEPH** | **DELPHI** | **L3** | **OPAL** | **ATLAS** | **CMS** | **ILD** | **SiD** |
|---|---|---|---|---|---|---|---|---|
| **B** (T) | 1.5 | 1.2 | 0.5 | 0.435 | 2 | 4 | 3.5 | 5 |
| **R** (m) | 2.68 | 2.6 | 5.9 | 2.2 | 1.25 | 2.95 | 3.6 | 2.6 |
| **L** (m) | 7 | 6.8 | 11.9 | 6.3 | 5.8 | 13 | 7.3 | 6 |
| α | 2.6 | 2.6 | 2.0 | 2.9 | 4.6 | 4.4 | 2.0 | 2.3 |
| cos θ | 0.79 | 0.79 | 0.5 | 0.82 | 0.92 | 0.91 | 0.5 | 0.75 |
| **U** (GJ) | 0.137 | 0.108 | 0.150 | - | 0.04 | 2.7 | 2.3 | 1.6 |

Table 1: Coil parameters for LEP, LHC, ILC experiments

The aspect parameter for electron colliders is roughly centred at 2.3-2.6, whilst being about 4.5 for LHC.
The angular coverage varies from 0.5 (L3, ILD) to 0.8-0.9 (other experiments).
The stored energy is one order of magnitude higher for LHC and ILC than for LEP experiments, CMS reaching the limit of 2.7 GJ.

### 2.2 The Vertex Detector

The physics motivation for a high-precision vertex detector is to provide good pattern recognition capability for charged tracking and excellent impact parameter resolution, allowing flavour identification of jets, by excellent reconstruction of separated vertices and of the tracks associated to them. To meet these requirements, a low-mass pixel detector as close as the interaction point as possible is a must.

Uncertainty on the transverse impact parameter depends on the radii of the different layers of the vertex detector, and on the space point precision as measured in each layer. The angular dispersion $\theta_0$ resulting from multiple scattering is inversely proportional to the track momentum p and depends on the material thickness x ( $X_0$ is the radiation length of the considered material) accordingly to the relation:

$$\theta_0 = (13.6 / \beta cp ) z \sqrt{x/X_0} \; [ 1 + 0.038 \ln (x/X_0) ]$$

The best precision is obtained from small radius, minimum thickness, and large momentum.

All four LEP experiments were equipped with vertex detectors based on silicon sensors arranged in 2 or 3 (DELPHI) layers of single (OPAL) or double sided microstrips, corresponding to a total thickness of about 1.5% of $X_0$ (2 layers) at normal incidence, and achieving similar impact parameter resolution of typically 100 μm (rϕ) and 150 μm (z) at 1 GeV, and about 30 μm at 20 GeV.

The LHC experiments have implemented silicon pixels in 3 layers, of roughly 10% of $X_0$ total thickness at normal incidence. ATLAS and CMS achieve typical resolutions of 100 μm at 1 GeV and 20 μm at 20 GeV.

The ILC experiments consider a vertex detector with 5 to 6 layers of silicon-based sensors, not exceeding 1% $X_0$ total thickness, and capable to reach a transverse impact-parameter resolution of 10 μm at 1 GeV and 2 μm at 20 GeV.

Table 2 summarizes the main characteristics of the LEP, LHC and ILC vertex detectors: number of layers, radius of each layer, type of sensor, amount of material (at 90º), pixel size, single point resolution, impact parameter resolution (for particles of typically 10 GeV transverse momentum).

## 2.3 The Main Tracker

The main tracker combined with the inner tracker aims at measuring the trajectory of charged particles in the tracking volume from several points measured along the track. A curve is fit to the points and from the curvature of the trajectory in the magnetic field the momentum of the charged track is measured.

Precise measurements necessitate large volumes, high fields and precise space-point measurements. In addition best performance of the Particle-Flow Algorithms require fine readout granularity for pattern recognition.

Two main options have been chosen by the LEP, LHC, and ILC experiments:

- drift chambers of different types:
    - time-projection chamber (TPC) for ALEPH, DELPHI, ILD
    - time-expansion chamber (TEC) for L3
    - Jet chamber (JC) for OPAL
- Silicon strips for ATLAS (combined with straw tubes), CMS, SiD

Table 3 summarizes the main characteristics of these detectors: type, number of layers (for Si-based detectors), inner and outer radius, length, material budget, point resolution, momentum resolution.

The best performances are achieved through silicon-strip sensors with the drawback of more material in general, and higher cost for large detectors.

|  | ALEPH | DELPHI | L3 | OPAL | ATLAS | CMS | ILD | SiD |
|---|---|---|---|---|---|---|---|---|
| **Layers** | 2 | 3 | 2 | 2 | 3 | 3 | 3 | 5 |
| **Radii (cm)** | 63<br>110 | 66<br>92<br>106 | 64<br>79 | 61<br>74 | 50.5<br>88.5<br>122.5 | 44<br>73<br>102 | 16<br>37<br>58 | 14<br>22<br>35<br>48<br>60 |
| **Sensor type** | Double-sided | Double + single | Double-sided | Single-sided | Single-sided | Single-sided | Double-sided | Single-sided |
| **Material (% $X_0$)** | 1.5 | 3.1 | 1.2 | 1.5 | 10 | 10 | 0.9 | 0.5 |
| **Pixel size (r$\phi$ x z) ($\mu m^2$)** | - | - | - | - | 40 x 400 | 100 x 150 | 10 x 10 | 20 x 20 |
| **Point resolution (R$\phi$) ($\mu m$)** | 10 | 8 | 8 | 8-10 | 10 | 15-20 | 3 | 6 |
| **Point resolution (z) ($\mu m$)** | 15 | 11 | 20 | 10-12 | 115 | 15-20 | 3 | 6 |
| **Impact param. resol. (R$\phi$) ($\mu m$)** | 34 | 25 | 30 | 18 | 20 | 20 | 2 | 2 |
| **Impact param. resol. (z) ($\mu m$)** | 34 | 34 | 130 | 24 | 50-100 | 20-30 | 2 | 2 |

Table 2: vertex-detector main parameters

|  | **ALEPH** | **DELPHI** | **L3** | **OPAL** | **ATLAS** | **CMS** | **ILD** | **SiD** |
|---|---|---|---|---|---|---|---|---|
| **Type** | TPC | TPC | TEC | JC | Si strips Straws | Si strips | Si strips TPC Si strips | Si strips |
| **Layers** | - | - | - | - | 4 x2 (Si) 36 (st.) | 10 | - | 5 |
| **Rin**(cm) | 31 | 29 | 17 | 25 | 30 (Si) 56 (st.) | 20 | 33 | 22 |
| **Rout**(cm) | 180 | 122 | 94 | 183 | 52 (Si) 107 (st.) | 116 | 181 | 122 |
| **Length** (cm) | 470 | 260 | 126 | 400 | 150 | 540 | 470 | 111-304 |
| **Material** (% $X_0$) | 7.1 | - | 7 | 4 | 1.2 10 | 30 | 5 | 10-15 |
| **Point resolution** (Rφ) (μm) | 150 | 250 | 50 | 120 | 17 170 | 15 | 60-100 | 8 |
| **TPC only** $\sigma(1/p_T)$ (/GeV) | $1.2 \times 10^{-3}$ | $5.0 \times 10^{-3}$ | - | - | - | - | $9 \times 10^{-5}$ | - |
| **Global** $\sigma(1/p_T)$ (/GeV) | $6 \times 10^{-4}$ | $6 \times 10^{-4}$ | $2.1 \times 10^{-2}$ | $1.5 \times 10^{-3}$ | $3.5 \times 10^{-4}$ | $1.5 \times 10^{-4}$ | $2 \times 10^{-5}$ | $2-5 \times 10^{-5}$ |

Table 3: Characteristics of main trackers (CMS resolution is the global resolution)

In the physics environment of ILC, a TPC has been considered for the ILD detector. The pros are the large experience acquired with this technology, the possibility of measuring tracks with a large number of three-dimensional space points, the continuous tracking, the easy reconstruction, the minimal amount of material in the tracking volume, the bonus of particle identification through dE/dx measurement.

The cons are the ion backflow, the possible overlap of multiple events, the moderate precision on space-point resolution and double-hit resolution (compensated by continuous tracking), the increase of size and cost of the calorimeters and solenoid.

## 2.4 The Calorimeters

Particle-Flow Algorithms (PFA) have been successfully applied since the LEP era to many detectors. ALEPH [12], an even CMS [13], in more difficult conditions, have demonstrated significant improvements of the jet energy resolution compared to methods based on calorimetric measurements alone. None of these detectors had been designed to make optimal use of the PFA method. This method is based on the jet energy resolution and relies heavily on the correct assignment of energy cluster deposits to the charged or neutral particles, depending on the transverse and longitudinal granularity of the calorimeters and on their resolution.

Table 4 and 5 display the main characteristics of LEP, LHC and ILC detectors respectively: absorbing material, sensitive medium, number of radiation length $X_0$ in ECAL, number of absorption length Λ in HCAL, resolution (stochastic term a, noise term b, constant term c); resolutions at 50, 150 and 500 GeV are also given in order to ease the comparison.

|  | ALEPH | DELPHI | L3 | OPAL | ATLAS | CMS | ILD | SiD |
|---|---|---|---|---|---|---|---|---|
| **Absorber** | Pb | Pb | BGO | Lead glass | Pb | PbWO4 | W | W |
| **Detector** | Wire chamber | HPC | BGO | Lead glass | Liq.Ar | PbWO4 | Si or Sc. | Si |
| **$X_0$** | 22 (4,9,9) | 18 (9 samp) | 22 | 24.6 | 25 (6,16,3) | 25 | 24 | 26 |
| **Granul.** | $0.8^0$ | $0.5^0$ | $2.3^0$ | $2.3^0$ | $1.2^0$ | $1^0$ | $0.25^0$ | $0.2^0$ |
| **σE/E a** | 0.18 | 0.32 | 0.02 | 0.15 | 0.10 | 0.03 | 0.17 | 0.17 |
| **σE/E b** | - | - | - | - | - | 0.25 | - | - |
| **σE/E c** | 0.009 | 0.043 | 0.005 | 0.002 | 0.02 | 0.006 | 0.01 | 0.01 |
| **σE/E (%) @50 GeV** | 2.7 | 6.2 | 0.6 | 2.1 | 2.5 | 0.9 | 2.6 | 2.6 |
| **σE/E (%) @150 GeV** | 1.7 | 5.0 | 0.5 | 1.2 | 2.2 | 0.7 | 1.7 | 1.7 |
| **σE/E (%) @500 GeV** | 1.2 | 4.5 | 0.5 | 0.7 | 2.1 | 0.6 | 1.3 | 1.3 |

Table 4: Characteristics of ECAL calorimeters

|  | ALEPH | DELPHI | L3 | OPAL | ATLAS | CMS | ILD | SiD |
|---|---|---|---|---|---|---|---|---|
| **Absorber** | Fe | Fe | U | Fe | Fe | Brass | Steel | Steel |
| **Detector** | Stream. tubes | Stream. tubes | PWC | Stream. tubes | Sc. | Sc. | Sc. or RPC | RPC |
| **Λ** | 7.16 | 6.6 | 3.36 | 4.8 | 7.2 | 5.8 | 5.5 | 4.5 |
| **Granul.** | $3.7^0$ | $3.0^0 \times 3.7^0$ | $2.5^0$ | $7.5^0$ | $5^0$ | $4^0$ | $1\text{-}2^0$ | $0.5^0$ |
| **σE/E a** | 0.85 | 1.12 | 0.55 | 1.2 | 0.52 | 1. | 0.5 | 0.6 |
| **σE/E b** | - | - | - | - | 1.6 | - | - | - |
| **σE/E c** | - | 0.21 | 0.05 | - | 0.03 | 0.05 | - | 0.08 |
| **σE/E (%) @50 GeV** | 12 | 26 | 9 | 17 | 9 | 11 | 7 | 12 |
| **σE/E (%) @150 GeV** | 7 | 23 | 7 | 10 | 5 | 10 | 4 | 9 |
| **σE/E (%) @500 GeV** | 4 | 22 | 6 | 5 | 4 | 7 | 2 | 8 |

Table 5: Characteristics of HCAL calorimeters

### 2.4.1 ECAL

Electromagnetic calorimeters are usually placed inside the solenoid, but in the case of OPAL and ATLAS.

OPAL had chosen a high energy-resolution calorimeter made of lead-glass Cerenkov counters readout with photomutipliers tubes, which have to be operated in a low magnetic field. This requirement imposed to place the electromagnetic calorimeter out of the solenoid, and in spite of the information provided by the pre-sampling device placed in front, the stochastic term is degraded from 0.06 to 0.15.

Although the electromagnetic calorimeter of ATLAS is placed outside of the solenoid, the thickness of the coil, combined with a sampling detector preceding the calorimeter, compensate the energy loss in the coil.

Electromagnetic calorimeters can be classified in two types:

- homogeneous calorimeters (L3, OPAL, CMS), for which the absorber is also the sensitive medium.

- sampling calorimeters (ALEPH, DELPHI, ATLAS, ILD, SiD) which are sandwiches made of alternate layers of absorber and sensitive media. The absorber material is mainly lead (LEP, LHC), or tungsten (ILC). The sensitive medium is either a gaseous detector (LEP, ILC) or a silicon-based detector (ILC), a scintillator-based detector (ILD).

The depth of these calorimeters ranges from 22 to 26 $X_0$ (about 1 $\Lambda$), with granularities spreading from $0.2^0$–$0.5^0$ (ILC, L3) to $0.8^0$-$1^0$ (ALEPH, LHC) and $2.3^0$ (LEP, but L3).

In terms of energy resolution, the experiments can grossly be separated in two groups: homogeneous calorimeters (L3, OPAL, CMS) with a stochastic term ranging from 0.02 to 0.06, and the sampling calorimeters (ALEPH, DELPHI, ATLAS, ILC) with stochastic terms ranging from 0.15 to 0.18.

### 2.4.2 HCAL

Placing the HCAL inside or outside the coil is questionable. Three (ALEPH, DELPHI, OPAL) out of the four LEP experiments and ATLAS have chosen to place the hadron calorimeter outside the solenoid using the return iron yoke of the coil as a natural solution for the absorber material. CMS as well as ILC experiments have preferred placing the hadron calorimeter inside the solenoid; the calorimeter experiencing a strong magnetic field is necessarily made of non-magnetic material (copper alloy or stainless steel). The active material is either a gaseous detector or a scintillator detector.

The depth of these calorimeters ranges from 3.4 $\Lambda$ (L3) to 7 $\Lambda$ (ALEPH, ATLAS), with a middle value of 4.5-5.5 $\Lambda$ (OPAL, CMS, ILC). A wide range of granularities has been selected, from $0.5^0$–$2^0$ (ILC) to $2.5^0$ (L3), $4^0$–$5^0$ (LHC), $7.5^0$ (OPAL).

In terms of energy resolution, the stochastic term groups around two values: 0.5-0.6 (L3, ATLAS, ILC), and 1-1.2 (DELPHI, OPAL, CMS); ALEPH stochastic term lies in the middle, reaching a resolution comparable to that of ILC experiments above 50 GeV.

### 2.5 The Muon Detector

The muon detector is designed to identify muons and to reject hadrons with high efficiency. The information from this detector, combined with that of other tracking devices allows 3-dimensional track reconstruction through the entire detector for each muon candidate.

Table 6 displays the main characteristics of the muon detectors at LEP, LHC, and ILC: technology, number of layers and spacing between them, position resolution, identification efficiency and background (dependent on loose or tight selection, 90% is typically obtained for 1% background).

When the return yoke of the solenoid has not been used for hadron calorimetry, the muon detectors are installed in the gaps between iron layers of the solenoid flux return. In other cases, but ATLAS, the muon detectors are essentially surrounding the hadron calorimeter (ALEPH, OPAL, DELPHI), or inside the coil (L3). In the ATLAS experiment, the muon chambers are placed between and on the coils of the superconducting barrel toroid magnet.

Muon detectors cover large areas, and reliability and low cost requirement drive the choice of gaseous detectors (stream tubes, drift chambers, monitored drift tubes, resistive plate chambers) in most cases, although scintillators are a possible option for ILC experiments.

|  | **ALEPH** | **DELPHI** | **L3** | **OPAL** | **ATLAS** | **CMS** | **ILD** | **SiD** |
|---|---|---|---|---|---|---|---|---|
| **Type** | Stream tubes | Drift chamber | Drift chamber | Drift chamber | MDT | DT, RPC | Sc/RPC | Sc/ RPC |
| **Layers** | 2 double | 4 | 5 | 4 | 3 | 4 | 10 + 3 | 10 |
| **Spacing (cm)** | 50 | 20 | - | - | 250 | - | 14 (10 l) 60 (3l) | - |
| **Resol. Pos. R$\phi$ (mm)** | - | 1.5 | 0.2 | 1.5 | - | 0.1 | - | - |
| **Resol. Pos. z (mm)** | 1 | 2 | 0.5 | 2 | 0.035 | 0.15 | a few | 2 |
| **Resol. dir. (mrad)** | 3.5 | 5 | 1 | - | - | - | - | - |
| **Efficiency (%)** | 86 | 66-96 | 90 | 85 | 80 | > 95 | 97 | 90 |
| **Fake $\mu$ (%)** | 0.8 | 0.5-8 | 0.7 | 1.3 | < 2 | 0.1-1 | a few | - |

Table 6: Characteristics of Muon detectors

## 3. Benchmark Physics Processes

Physics performance studies are necessary to quantify the physics performance of a future detector for FCC-ee. A set of benchmark physics processes places the most stringent requirements on the detector parameters.

Accurate measurements of branching fractions of a light Higgs boson to cc, or bb, $\tau$ identification are some measurements that require precise vertexing capability.

The tracker momentum resolution is tested by the requirements of the recoil mass analysis of the Higgs-strahung process $e^+e^- \to Zh \to l^+l^-X$, for instance.

The need to distinguish hadronically decaying W, Z and h bosons from one another in processes like $e^+e^- \to W^+W^-$, ZZ, Zh, Zhh, drives the FCC-ee jet-energy resolution requirements to a level that can only be reached with the use of PF algorithms combining information from calorimeters and trackers. This leads to strong demands on the calorimeter parameters, not only on resolution, but also on transversal and longitudinal granularity.

These are just examples, and a list of selected benchmark physics processes needs to be established to classify the various proposed detectors according to their figure of merit.

This process requires a great effort of simulation and analysis, and at this stage the present proposal is only based on the comparison of LEP experiments, and LHC experiments in a few physics channels.

### 3.1 LEP experiments

Table 7 displays the measurements of the W mass and width in the four LEP experiments ([14] – [17]). Systematic errors are summarized in Tables 8 and 9.

The analysis of these tables shows that in the electron channel, a high resolution such as that of the L3 electromagnetic calorimeter, does not necessarily result in the smallest total systematic error from calorimetry. Actually ALEPH reaches almost the same level of precision due to the good mastering of the energy scale. Almost all four experiments obtain similar uncertainties in the muon channel. In the tau channel as in the purely hadronic channel, ALEPH is favoured by the high jet performance enhanced by the use of the PFA algorithm as well as the good particle identification.

| | ALEPH | DELPHI | L3 | OPAL |
|---|---|---|---|---|
| $m_W^{eq}$(GeV) | 80.536±0.087±0.027 | 80.388±0.133±0.036 | 80.225±0.099±0.024 | - |
| $m_W^{\mu q}$(GeV) | 80.353±0.082±0.025 | 80.294±0.098±0.028 | 80.152±0.119±0.024 | - |
| $m_W^{\tau q}$(GeV) | 80.394±0.121±0.031 | 80.387±0.144±0.033 | 80.195±0.175±0.060 | - |
| $m_W^{lq}$(GeV) | 80.429±0.054±0.025 | 80.339±0.069±0.029 | 80.196±0.070±0.026 | 80.449±0.056±0.028 |
| $m_W^{qq}$(GeV) | 80.475±0.070±0.028 ±0.028 (FSI) | 80.311±0.059±0.032 ±0.119 (FSI) | 80.298±0.064±0.049 (FSI incl.) | 80.353±0.060±0.058 (FSI incl.) |
| $\Gamma_W^{eq}$(GeV) | 1.84±0.20±0.08 | - | - | - |
| $\Gamma_W^{\mu q}$(GeV) | 2.17±0.20±0.06 | - | - | - |
| $\Gamma_W^{\tau q}$(GeV) | 2.01±0.32±0.06 | - | - | - |
| $\Gamma_W^{lq}$(GeV) | 2.01±0.13±0.06 | 2.452±0.184±0.073 | - | 1.927±0.135±0.091 |
| $\Gamma_W^{qq}$(GeV) | 2.31±0.12±0.04 ±0.11 (FSI) | 2.237±0.137±0.139 ±0.0248 (FSI) | 1.97±0.11±0.09 | 2.125±0.112±0.177 |

Table 7: Results on $m_W$ and $\Gamma_W$ in the eνqq, μνqq, τνqq, lvqq, qqqq channels. The first uncertainty is statistical, the second uncertainty is systematic.

| $m_W$(MeV) | ALEPH | DELPHI | L3 | OPAL |
|---|---|---|---|---|
| l En scale | 3 / 8 / - / - | 25 / 21 / - / - | 6 / 12 / - / - | 2 / 8 / - / - |
| l En resol | 12 / 4 / - / - | 15 / 10 / 17/ 4 | | 2 / 2 / - / - |
| j En scale | 5 / 5 / 9 / 2 | 11 / 9 / 16 / 8 | 4 / 11/ 23 / 5 | 7 / 4 |
| j En resol | 4 / 2 / 8 / - | 8 / 5 / 8 / 10 | | 1 / 0 |
| j ang. bias | 5 / 5 / 4 / 6 | 3 / 5 / 5 / 2 | | 4 / 7 |
| j ang. resol | 3 / 2 / 3 / 1 | - / - / - / 1 | | 0 / 0 |
| Hadronisation | 20 / 20 / 25 / 17 | 10 / 10 / 13 /12 | 11 / 12 / 44 / 20 | 14 / 20 |
| Rad. cor. | 3 / 2 / 3 / 2 | 9 / 4 / 5 / 5 | 16 / 10 / 9 / 6 | 11 / 9 |
| LEP beam en. | 9 / 9 / 10 / 9 | 15 / 15 / 15 / 15 | 10 / 10 / 10 / 10 | 8 / 10 |
| Color recon. | - / - / - / 79 | - / - / - / 212 | - / - / - / 38 | - / 41 |
| BE corel. | - / - / - / 6 | - / - / - / 31 | - / - / - / 17 | - / 19 |

Table 8: Summary of the systematic errors on $m_W$ in the eνqq, μνqq, τνqq, and qqqq channels. The three numbers in each cell correspond to these four channels. For OPAL, some numbers are not available and then only lvqq, and qqqq are quoted.

| $\Gamma_W$(MeV) | ALEPH | DELPHI | L3 | OPAL |
|---|---|---|---|---|
| l En scale | 5 / 4 / - / - | 48 / - | 12 / 37 / - / - | 7 / 8 / - / - |
| l En resol | 65 / 55 / - / - | 15 / 9 | | 27 / - / - / - |
| j En scale | 4 / 4 / 16 / 2 | 38 / 169 | 20 / 30 / 75 / 20 | 0 / 0 |
| j En resol | 10 / 18 / 36 / 7 | | | 16 / 4 |
| j ang. bias | 2 / 2 / 3 / 1 | | | 2 / 0 |
| j ang. resol | 6 / 7 / 8 / 15 | | | 2 / 4 |
| Hadronisation | 22 / 23 / 37 / 20 | 29 / 8 | 55 / 70 / 150 / 85 | 77 / 68 |
| Rad. cor. | 3 / 2 / 2 / 5 | 11 / 9 | 5 / 5 / 5 / 5 | 11 / 10 |
| LEP beam en. | 7 / 7 / 10 / 7 | 15 / 9 | 5 / 5 / 5 / 5 | 3 / 2 |
| Color recon. | - / - / - / 104 | - / 247 | - / - / - / 50 | - / - / - / 151 |
| BE corel. | - / - / - / 20 | - / 20 | - / - / - / 10 | - / - / - / 32 |

Table 9: Summary of the systematic errors on $\Gamma_W$ in the eνqq, μνqq, τνqq, and qqqq channels. The three numbers in each cell correspond to these four channels. For OPAL, some numbers are not available and then only lvqq, and qqqq are quoted.

## 3.1 LHC experiments

Table 10 shows the measurements of the top mass ([18]) and Higgs mass ([19]) in ATLAS and CMS experiments. Systematic errors on $m_t$ are summarized in Table 11.

The examination of the results for $m_t$ reveals similar performances in the di-lepton channel, with a slightly better systematic error for CMS that takes advantage of the high energy-resolution of the electromagnetic calorimeter. In the semi-leptonic and purely hadronic channels, CMS shows better performances in the jet reconstruction, in spite of the poor energy resolution of the hadron calorimetry.

|  | ATLAS | CMS |
|---|---|---|
| $m_t^{ll}$(GeV) | 173.09 ± 0.64 ± 1.50 | 172.50 ± 0.43 ± 1.46 |
| $m_t^{lj}$(GeV) | 172.31 ± 0.23 ± 1.35 ± 0.72 (JES) | 173.49 ± 0.27 ± 0.98 ± 0.33 (JES) |
| $m_t^{jj}$(GeV) | 174.9 ± 2.1 ± 3.8 | 173.49 ± 0.69 ± 1.23 |
| $m_H^{\gamma\gamma}$(GeV) | 126.8 ± 0.2 ± 0.7 (125.98 ± 0.42 ± 0.28) | 125.4 ± 0.5 ± 0.6 |
| $m_H^{4l}$(GeV) | 124.3 ± 0.5 ± 0.5 (124.51 ± 0.52 ± 0.06) | 125.8 ± 0.5 ± 0.2 (125.6 ± 0.4 ± 0.2) |

Table 10: Results on $m_H$[3] in the γγ and four-lepton channels, and $m_t$ in the dilepton, l+jets, all jets channels. The first uncertainty is statistical, the second uncertainty is systematic.

| $m_t$(GeV) | ATLAS | CMS |
|---|---|---|
| j En. scale | 0.88 / 1.07 / 2.1 | 0.97 / 0.42 / 0.97 |
| b-jet En. scale | 0.71 / 0.08 / 1.4 | 0.76 / 0.61 / 0.49 |
| j En. resol. | 0.21 / 0.22 / 0.3 | 0.14 / 0.23 / 0.15 |
| j reco eff. | - / 0.05 / 0.2 | - |
| Method | 0.07 / 0.13 / 1.0 | 0.40 / 0.06 / 0.13 |
| MC gen | 0.20 / 0.19 / 0.5 | 0.04 / 0.02 / 0.19 |
| ISR / FSR | 0.37 / 0.45 / 1.7 | 0.58 / 0.30 / 0.32 |
| PDF | 0.12 / 0.17 / 0.6 | 0.09 / 0.07 / 0.06 |
| Backgd model. | 0.14 / 0.10 / 1.9 | 0.05 / 0.13 / 0.13 |

Table 11: Systematic uncertainty contributions on the measurement of $m_t$. The three numbers in each cell correspond to the dilepton, l+jets, all jets channels.

Two measurements of $m_H$ are given. The first one is that published at the time of the combination of the results of the two experiments, and similar precisions were obtained, with a slight advantage for CMS. In parentheses are the latest published numbers. CMS numbers have little changes, but spectacular and surprising improvement shows up in ATLAS systematic uncertainties.

## 4. The ASAHEL detector

In the light of what has been learned from the characteristics and performances of the LEP and LHC detectors, amended by the more recent studies done for the ILC experiments, and taking into account the differences between a circular collider and a linear collider, an ALEPH-like design is proposed for the ASAHEL detector. The magnet dimensions and field as well as the size and granularity of the individual sub-detectors will be optimized starting from this design.

---

[3] For $m_H$, masses are given at the time of the combination of ATLAS and CMS results. In parentheses are shown the individual updates.

The results obtained by the LEP and LHC experiments demonstrate that the precision of measurements suffer more from energy calibration and scale than from intrinsic energy resolution. They have stressed the huge improvement brought by the PFA algorithms that benefit from excellent tracking capabilities and high granularity calorimetry: a jet energy resolution of $0.5/\sqrt{E_J}$ that is about a factor two better than that obtained from a calorimetric approach can be obtained.

At the FCC-ee , as at the ILC, physics is characterized by high multiplicity final states with multijets. With such conditions jet resolution is crucial for physics sensitivity. The PFA has demonstrated to be a powerful tool to measure jet energy with the required accuracy, exploiting the fact that most of the jet energy is carried by charged particles which particle momentum is better measured in the tracker than in the calorimeters, and then separating and reconstructing particles using tracker for charged particles, ECAL for photons, ECAL+HCAL for neutral hadrons. The intrinsic energy resolution (single particles) is not a limitation for the jet energy resolution, to which the largest contribution comes from wrong assignment of energy deposits to reconstructed particles. To minimize this latter contribution, a large detector to separate particles, a high magnetic field to separate charged from neutrals, a high calorimetric granularity to resolve energy deposits, are desirable, but leading to cost increase, these parameters need to be carefully optimized through simulation work. A possible figure of merit, including these 3 aspects, might be defined as the ratio $BR^2/R_M^{eff}$, where R is the inner radius of the calorimeter and $R_M^{eff}$ the effective Molière radius[4]. A comparison of this figure for ALEPH and ILC detectors is shown in table 17 of section 4.4.1.

Another important quality is particle identification associated to good vertexing capability.

The ALEPH philosophy based on using as few detection techniques as possible proved to be rewarding and this is the decision made for ASAHEL. With all these considerations the proposed detector grossely follows ALEPH design with some modifications adapted to the FCC-ee energy range. Only the barrel part of the detector is described, as, according to ALEPH philosophy, the design criteria for the end-caps are the same as for the barrel, with some adaptations for different geometries.

Figure 1 shows a view of the ALEPH detector along the beam line. The choice of a solenoid is retained with ALEPH dimensions and a field of 3.5-5 Tesla (very similar to ILC). The vertex detector is installed as close as possible to the beam pipe and is surrounded by the central tracker, a large time projection chamber (TPC). The electromagnetic calorimeter is situated inside the coil to reduce the amount of material the particles traverse before they enter the calorimeter. In the ALEPH design, the hadron calorimeter, outside the coil, serves as the magnetic field return yoke. However for FCC-ee this option has to be evaluated with respect to the option inside the coil, as this latter configuration eases the track to cluster association for optimized PFA. These two possible choices necessarily impact the absorber depth and material. They have also consequences on the setup of muon chambers that can use the mechanical support provided by the yoke when the hadron calorimeter is placed inside the coil. These detectors are described in more details in the next subsections.

---

[4] $R_M^{eff} = f\ R_M$ , where $f = (x_a + x_s) / x_a$; $x_a$ ($x_s$) is the absorber (sensor) thickness

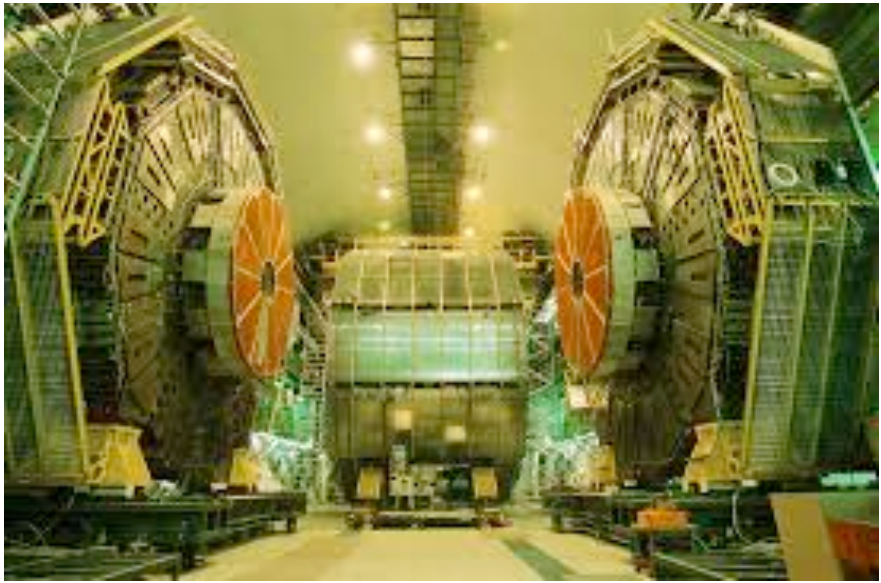

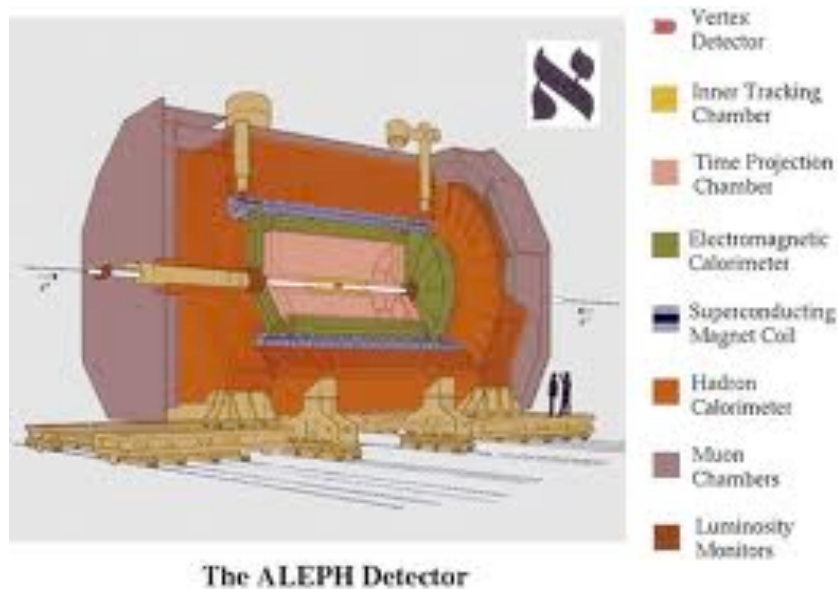

Figure 1: The ALEPH detector

### 4.1 The Magnet

The proposed solenoid is that of the SiD which parameters are summarized in Table 12. However the field may be tuned from 3.5 to 5 T.
The choice of the magnet basically obeys the same guidelines as those that had led to the design of the spectrometer magnets for LEP. Choosing the best device for a particular detector is an iterative process, starting with the comparison in terms of their resolving power based on measurements of sagitta, drawing attention to the importance of detector size. It is important to precisely define the expected resolution, and not to demand from the device more resolution than necessary, to avoid unnecessary increase of complications and cost. Thus starting from the basic design, the parameters will be optimized, considering the following elements:

- the expected spectrum of charged particles coming from an interaction
- multiple scattering in the beam pipe and in the detector that limits the possible momentum resolution

- in the case of a solenoid of radius R producing a field B, spiralling of charged particles with momentum p < 0.3 BR within the track detector.

|  | SiD | CMS |
|---|---|---|
| **Central Field** (T) | 5.0 | 4.0 |
| **Stored Energy** (GJ) | 1.59 | 2.69 |
| **Stored Energy Per Unit Cold Mass** ( kJ/kg) | 12 | 11.6 |
| **Operating Current** (kA) | 17.724 | 19.2 |
| **Inductance** (H) | 9.9 | 14.2 |
| **Fast discharge Voltage to Ground** (V) | 300 | 300 |
| **Number of layers** | 6 | 4 |
| **Total Number of Turns** | 1459 | 2168 |
| **Peak Field on Superconductor** (T) | 5.75 | 4.6 |
| **Temperature Stability Margin** (K) | 1.6 | 1.8 |
| **Total Cold Mass of Solenoid** | 130 | 220 |
| **Number of Winding Modules** | 2 | 5 |
| $R_{min}$ **Cryostat** (m) | 2.591 | 2.97 |
| $R_{min}$ **Coil** (m) | 2.731 | 3.18 |
| $R_{max}$ **Coil** (m) | 3.112 | 3.49 |
| $R_{max}$ **Cryostat** (m) | 3.392 | 3.78 |
| $Z_{mmax}$ **Cryostat** (m) | ± 3.033 | ± 6.5 |
| $Z_{mmax}$ **Coil** (m) | ± 2.793 | ± 3.2 |
| **Operating Temperature** (K) | 4.5 | 4.5 |
| **Cooling Method** | Forced flow | Termosiphon |

Table 12: SiD and CMS Superconducting Coil Comparison

### 4.2 The Vertex Detector

The identification of heavy quarks and tau leptons is essential for the FCC-ee physics program as it is for that of ILC. The reconstruction of decay vertices of short-lived particles requires a particularly light and precise vertex detector able to measure the charged particle track parameters with great accuracy that can be quantified through the resolution on the impact parameter of charged particles. To reach this goal, the ILC experiments consider that the vertex detector should comply with the following specifications:

- Hit resolution better than 3-5 μ
- Less than 0.1-0.3 % radiation length per layer

Both ILC experiments have chosen pixel detectors. The hit resolution is directly related to the pixel size, which then varies typically from 10 to 20 μm. Larger pixel sizes can be used if charge sharing is used to improve the resolution. The bunch structure of ILC, with a 1 ms bunch train at 5 Hz enables power pulsing of the electronics, providing a power saving of a factor 50-1000 for front-end analog power. At FCC-ee power dissipation and cooling might be challenging, as the large repetition rate does not allow power pulsing. All these considerations have to be taken into account for the optimization of the pixel size and for the choice of the technology. The starting point for this could be the SiD design summarized in Table 14. Each layer is assumed to be 0.1 radiation length thick.

### 4.3 The Central tracker

The unique pattern recognition capabilities of the TPC, with the bonus of particle identification, makes it a relevant choice for FCC-ee, as well as for ILC, with some caveats that are presently studied. This is the option retained by the ILD Collaboration. The TPC is complemented by a silicon envelope made of two barrel components, the Silicon Inner Tracker (SIT) and the Silicon External Tracker (SET), one end cap component behind the endplate of the TPC (ETD), and the forward tracker (FTD).

| BARREL | R (mm) | $Z_{max}$ (mm) | |
|---|---|---|---|
| Layer 1 | 14 | 63 | |
| Layer 2 | 22 | 63 | |
| Layer 3 | 35 | 63 | |
| Layer 4 | 48 | 63 | |
| Layer 5 | 60 | 63 | |
| DISK | $R_{inner}$ (mm) | $R_{outer}$ (mm) | $Z_{center}$ (mm) |
| Disk 1 | 14 | 71 | 72 |
| Disk 2 | 16 | 71 | 92 |
| Disk 3 | 18 | 71 | 123 |
| Disk 4 | 20 | 71 | 172 |
| FORWARD DISK | $R_{inner}$ (mm) | $R_{outer}$ (mm) | $Z_{center}$ (mm) |
| Disk 1 | 28 | 166 | 207 |
| Disk 2 | 76 | 166 | 541 |
| Disk 3 | 117 | 166 | 852 |

Table 14: The geometry parameters of the SiD vertex detector (barrel, disks and forward disks.

The barrel silicon parts SIT and SET provide precise space points before and after the TPC; this improves the overall momentum resolution, helps in linking the vertex detector with the TPC, and in extrapolating from the TPC to the calorimeter. The coverage of the TPC with silicon tracking is completed by the ETD, located within the gap separating the TPC and the end-cap calorimeter. Together these systems help in calibrating the overall tracking system, in particular the TPC.

The main parameters of the central silicon system are displayed in Table 15.

| SIT (baseline = false double-sided Si microstrips) | | | | | |
|---|---|---|---|---|---|
| Geometry | | | Characteristics | | Material |
| R (mm) | Z (mm) | cos θ | Resolution (μm) | Time (ns) | $X_0$ (%) |
| 153 | 368 | 0.910 | 7.0 | 307.7 | 0.65 |
| 300 | 644 | 0.902 | 50.0 | 307.7 | 0.65 |
| SET (baseline = false double-sided Si microstrips) | | | | | |
| Geometry | | | Characteristics | | |
| R (mm) | Z (mm) | cos θ | Resolution (μm) | Time (ns) | $X_0$ (%) |
| 1811 | 2350 | 0.789 | 7.0 | 307.7 | 0.65 |
| ETD (baseline = single-sided Si microstrips) | | | | | |
| Geometry | | | Characteristics | | |
| R (mm) | Z (mm) | cos θ | Resolution (μm) | | $X_0$ (%) |
| 419.3 - 1822.7 | 2420 | 0.985 – 0.799 | 7.0 | | 0.65 |

Table 15: Main parameters of the central silicon systems SIT, SET, and ETD.

This central silicon system could be adapted for FCC-ee, using the same pixel sensors as those for the vertex detector.

The LCTPC (Linear Collider TPC) collaboration pursues R&D to develop a TPC based on the best-state-of-the-art technology [20]. Precision physics measurements at ILC and at FCC-ee require a very accurate momentum resolution of $2 \times 10^{-5}$ / GeV/c of the whole tracking system, implying a resolution of $9 \times 10^{-5}$ for the TPC immersed in a magnetic field of 3.5 T. To achieve these specifications, the point resolution in the R-ϕ plane is required to be smaller than 100 μm.

For the amplification and readout of the TPC, micropattern gas detectors (MPGD) are used instead of traditional wire chambers (ALEPH). Several prototypes of modules using Micromegas or GEM structures for gas amplification have been built and tested. The design of the final TPC at the ILD has started with activities on the mechanics, readout electronics, cooling solutions, and integration. Small prototypes are used for the study of backgrounds, ion backflow and gating.

The TPC that might be chosen for a FCC-ee detector would fully benefit from these studies. IRFU has a group actively working on the TPC for ILC, and a group of FCC-ee in this lab has now joined them to investigate how the different machine conditions would affect the operation of the TPC, particularly how electric field distortions caused by positively charged ions would affect the position resolution at the highest luminosity envisaged at FCC-ee ($10^{36}$ cm$^{-2}$s$^{-1}$).

Simulations have been carried out in the following conditions [S. Ganjour, P. Schwemling, "Distorsions due to Ions in a TPC for FCC-ee", FCC-Saclay meeting, 6 Nov. 2014"]. A luminosity of 5.6 x $10^{34}$ cm$^{-2}$s$^{-1}$ at the Z pole has been assumed, with 16.8 kHz hadronic Z decays, 33.6 kHZ Bhabhas, ~ 0.3 kHz γγ interactions and the TPC parameters are those of ILD, with the exception of the inner cage that is placed at 400 mm instead of 229 mm. The charge distribution of ions stored in the TPC volume (primaries and backflow) has been estimated analytically, Pythia and ILC distortion-computation program have been used, and a Ion-Backflow Factor of 1 has been assumed. Primary ions are produced in the TPC drift region by ionisation of the gas by charged particles. The resulting distortions of the electric field produce O(1μm) position distortions. More dangerous are the secondary ions produced in the amplification region. Positive ions flowing from that region create a high-density ion disk for each train crossing, drifting back into the TPC volume. This disk slowly (1m/s) drifts to the cathode, affecting the electron drift in the subsequent bunch crossings. To prevent the ion feedback, gating devices can be placed in front of the amplification devices. Studies have been done for different magnetic field values: 0, 1.5, 3.5 T, and the maximum distortions was ~60 μm without gating. The estimated distorsions are compatible with those quoted for ILD, however there is little margin to keep a distorsion less than 100 μm at a luminosity of $10^{36}$ cm$^{-2}$s$^{-1}$.

The ion backflow can be reduced, by increasing the ratio between the amplification field and the drift field ($E_A/E_D$), and increasing the gap and pitch sizes. Micromegas with a high field ratio presents natural ion-backflow suppression, at the level of 1-2% for $E_A/E_D$ ~ 100; a factor 10 decrease would be desirable for safety margin.

### 4.4 The Calorimeters

High precision measurements require a jet energy resolution better than a few %, that impose a challenge on calorimetry in the association of energy deposits with either charged or neutral particles:

- enhanced separation of electrons and charged hadron tracks by minimising the lateral shower size of electromagnetic clusters with minimised Molière radius of the ECAL

- capability of assignment of energy cluster deposits to charged or neutral particles requiring fine transverse and longitudinal segmentation of both ECAL and HCAL.

- capability of doing optimal track to cluster association requiring ECAL and possibly HCAL to be inside the solenoid in order to avoid energy loss in the coil. Both options of HCAL inside or outside the coil are discussed.

- hermiticity requiring suitable calorimeter length to cover small angles and calorimeter depth to contain electromagnetic and hadronic showers.
- minimal gap between the tracker and the ECAL

To ease maintenance, monitoring and calibration, to concentrate expertise, the choice is made to use the same techniques in all places where it is possible. Hence, as Micromegas devices are considered for the TPC gas amplification and readout, this technique appears as a natural choice. Indeed using Micromegas chambers for hadron calorimetry has been studied for ILC [21]. To maximize the solid angle coverage (hermiticity), there must be as few cracks (dead regions) as possible and, where they inevitably occur, the calorimeters must be arranged so that cracks in one are covered by the others. This makes it unlikely that any neutral particle except a neutrino or some new non-interacting object could escape without leaving a mark.

The emphasis is placed on fine granularity with projective towers of the same size as the electromagnetic or hadronic showers.

### 4.4.2 ECAL

Table 4 in section 2 reveals that the ALEPH electromagnetic calorimeter had an energy resolution totally comparable to those for ILC, and a better granularity than the ECAL of CMS, that was used for the Physics Case of TLEP [1]. Thus the design for ASAHEL ECAL will be developed from ALEPH ECAL. This ECAL is a sampling calorimeter with alternate layers of lead and proportional tubes. There are 45 layers grouped in 3 stacks. The wire chambers are made of Al-extrusion, covered with graphite-coated Mylar. Signals are derived from the cathode pads and the wires. Cathode pads are arranged in projective towers, 49152 in the barrel and 24576 in the end-caps, read out in 3 storeys, The gas is Xe (80%) + $CO_2$ (20%). Dimensions and performances are shown in table 16.

| Dimensions and performances | | |
|---|---|---|
| **Al proportional tubes, active cross-section** | 3.2 x 4.5 $mm^2$ | |
| **Anode wires (gold-plated W), diameter** | 25 $\mu m$ | |
| **Average tower solid angle** | 1.67 x $10^{-4}$ sr | |
| **Operating gas pressure (atm. + 50 mbar)** | 990 mbar | |
| **Operating temperature, required precision** | $\pm 0.5^\circ$ C | |
| **Energy resolution $\sigma_E/E$** | 0.18 $GeV^{1/2}/\sqrt{E}$ | |
| **Position resolution $\sigma_x = \sigma_y$** | 6.8 mm $GeV^{1/2}/\sqrt{E}$ | |
| **Granularity at $\theta = 90^0$ (barrel)** | 17 x 17 $mrad^2$ | |
| **Granularity at $\theta = 45^0$ (barrel)** | 12 x 12 $mrad^2$ | |
| **Granularity at $\theta = 40^0$ (end-cap)** | 9 x 10 $mrad^2$ | |
| **Granularity at $\theta = 27^0$ (end-cap)** | 10 x 14 $mrad^2$ | |
| | **Barrel** | **End-Caps** |
| **Number of modules** | 12 | 12 x 2 |
| **Number of layers, stack 1 (2mm Pb)** | 10 (3.83 $X_0$) | 10 (3.46 $X_0$) |
| **Number of layers, stack 2 (2mm Pb)** | 23 (8.80 $X_0$) | 23 (8.86 $X_0$) |
| **Number of layers, stack 3 (4mm Pb)** | 12 (8.85 $X_0$) | 12 (8.91 $X_0$) |
| **Inner radius** (cm) | 184.7 | 54 |
| **Outer radius** (cm) | 225.4 | 234.8 |
| **Thickness** (cm) | 40.7 | 56.25 |
| **Weight per module** (tons) | 10.4 | 2.6 |
| **No. of anode wires** | 115404 | 118800 x 2 |
| **Wire length** (cm) | 463.0 | 8-174 |
| **Meas.uniform. of response over 1 module (5)** | 1.3-1.6 | 1-1.5 |
| **Angular coverage in $\theta$ (deg.)** | 40-140 | 13-41 |
| **Av. frac. of insensitivity** (% of solid angle) | 2 | 6 |

Table 16: Parameters of the electromagnetic calorimeter

The described geometry will be the starting point for ASAHEL design, whilst the replacement of wire chambers by Micromegas chambers will be investigated, Application of Micromegas for sampling calorimetry puts specific constraints on the design and performance of this gaseous detector. In particular, uniform and linear response, low noise and stability against high ionisation density deposits are prerequisites to achieving good energy resolution. Linear response property of Micromegas originates from a fast collection of avalanche ions at the mesh (25-100ns), a relatively low operating gas gain (a few thousands), and a small diffusion of avalanche electrons (~ 10-20 μm RMS). Anode patterns take many forms from simple microstrips to complex two-dimensional patterns. Advanced MGDPs are now reaching sizes of 1-2 m length at the prototype level [22].

Now arise the question of the absorber. ALEPH experiment used lead whilst both ILC experiments use tungsten. Table 17 shows some important characteristics of these absorbers in association with the sensors. The same quantities are displayed for the CMS homogeneous ECAL using PbWO4. In particular the figure of merit, as defined in section 4, for the ALEPH detector can be compared with those for ILC experiments. This comparison may give an idea of the role played by each of the three components, assuming a granularity comparable to the Molière radius. For example, ALEPH and ILD have comparable $BR^2$, but the contribution of $R_M^{eff}$ to the figure of merit is about 4 times that of ILD mainly due to the contribution of the sensor.

Although more expensive than lead, tungsten presents a number of advantages in terms or radiation length and Molière radius. High-energy showers may not be fully contained in the 22 radiation lengths of the ALEPH calorimeter, and four additional radiation lengths would increase the depth of the calorimeter by about 2.5 cm, and consequently the radius of the coil, in particular if the hadron calorimeter is also placed inside the coil. A smaller Molière radius means a better shower position and a better shower separation, due to smaller shower overlaps. The effective Molière radius also depends on the thickness of the gap between absorber plates, placing constraints on the sensor thickness. Both options will be studied.

In addition, the transverse and longitudinal granularity will be optimized for maximizing the performances while keeping minimal the number of readout channels.

| ECAL | ILC | ALEPH | CMS |
|---|---|---|---|
| **Absorber / Sensor** | W / Si | Pb / PWC | PbWO4 |
| **Density** (g/cm$^3$) | 19.3 | 11.35 | 8.28 |
| **Radiation length $X_0$** (cm) | 0.3504 | 0.5612 | 0.89 |
| **Depth** ($X_0$) | 26 | 22 (4+9+9) | 26 |
| **Layer (absorber + sensor) thickness** (cm) | W 20 x 0.25 + 10 x 0.5 Sensor 30 x 0.125 | Pb 33 x 0.2 + 12 x 0.4 Sensor 45 x 0.38 | 23 |
| **Absorber Molière radius** (cm) | 0.9327 | 1.602 | 1.959 |
| **$R_M^{eff}$** (effective Molière radius, cm) | 1.40 | 4.65 | 1.959 |
| **Interaction Length $\Lambda$** (cm) | 9.946 | 17.59 | 20.27 |
| **B** (T) | 3.5 (ILD) / 5 (SiD) | 1.5 | 4 |
| **R** (cm) | 125 | 185 | 129 |
| **$BR^2$** (T.m$^2$) | 5.5 (ILD) / 7.8 (SiD) | 5.1 | 6.6 |
| **$BR^2$ / $R_M^{eff}$** | 391 / 558 | 110 | 340 |
| **Depth** ($\Lambda$) | 0.7 | 0.9 | 1.1 |

Table 17: Comparison of absorbers for ILC and ALEPH

### 4.4.2 HCAL

A comparison of HCAL characteristics listed in Table 5 of section 2 shows that the energy resolution of the ALEPH HCAL is quite comparable to that expected for SiD HCAL, and better than that of CMS HCAL that was used for the Physics Case of TLEP [1]. The granularity is smaller than that of CMS HCAL and only a factor two larger than a possible option of ILD. ALEPH magnet iron is instrumented with 23 layers of plastic limited-streamer tubes, separated by iron sheets 5 cm thick. The tubes layers are equipped with pad readout, summed in towers for a localized measurement of the total deposited energy. Each tube is also coupled capacitatively to strips parallel to the wires. It is their hit pattern that is readout. Wire information is read out plane-by-plane (end-caps) and two planes by two planes (barrel) and used for the trigger. The parameters of the HCAL are displayed in Table 18.

| Dimensions and performances | |
|---|---|
| **Angular size of towers (larger in ϕ at small η)** | $3.7^0$ x $3.7^0$ |
| **Number of towers** | 4788 |
| **Total iron depth of a tower near θ = $90^0$, ϕ = $0^0$** | 120 cm |
| **Typical accuracy of energy measurement σE/E** | 0.84 $GeV^{-1/2}$ / $\sqrt{E}$ |
| **Typical spatial accuracy for strip coordinate measurement (perpendicular to strip direction)** | 0.35 cm |

Table 18: Parameters of the hadronic calorimeter

This geometry is proposed as a starting point for ASAHEL design with the replacement of streamer tubes by Micromegas chambers. Some tests have been carried out with Micromegas chambers for hadronic calorimetry at ILC [21], that the SiD collaboration considers as alternatives to RPCs. In this case the hadron calorimeter is placed outside the coil and the iron yoke is a natural absorber. The alternative of placing the calorimeter inside the coil will also be investigated and then the absorber is necessarily made of copper alloy or stainless steel. As for the ECAL, the granularity will be optimized.

### 4.5 The Muon Detector

In the perspective of using ALEPH design as a starting point for ASAHEL design, outside of the magnet, behind the last layer (10 cm) of the HCAL, two double-layers of streamer tubes (replaced by Micromegas chambers) would form the muon detector. Each single layer reads out two orthogonal coordinates using strips. Dimensions and performances of the ALEPH muon detector are listed in Table 19.

| Dimensions and performances | |
|---|---|
| **Total number of muon chambers** | 94 |
| **Distance between muon chamber double layers:** | |
|     Barrel and middle-angle chambers | 50 cm |
|     End-cap chambers | 40 cm |
| **Typical accuracy of muon exit angle measurement** | 10 mrad |
| **Typical misidentification probability** | |
|     To take a π for a μ | 0.7 % |
|     To take a K for a μ | 1.6 % |
| **Fraction of solid angle covered by the sensitive part of the muon chambers above $15^0$** | |
|     Inner layer | 92 % |
|     Outer layer | 85 % |

Table 19: Parameters of the muon detector

The digital signals from the streamer tubes of the HCAL are used for muon identification by tracking the progress of the particles through the calorimeter. Muons are identified by their penetration abilities with the background from penetrating hadron showers removed by pattern-recognition algorithms on the digital signals. The number of muon chamber layers can be adjusted depending on the position of the HCAL inside or outside the coil. Micromegas is one of the detector technologies that have been chosen for precision tracking and trigger purposes for the upgrade of the forward muon system of the ATLAS experiment [17].

### 4.5 The Luminosity detectors

Accurate knowledge of the luminosity is required, both the energy-integrated luminosity, as well as the luminosity as a function of energy, dL/dE. Low-angle Bhabha scattering detected by dedicated calorimeters can provide the necessary precision for the integrated luminosity. Acollinearity and energy measurements of Bhabha events can be used to extract dL/dE. In ALEPH, since 1992, the luminosity detectors were two tungsten-silicon calorimeters covering the 24-58 mrad angular interval. The characteristics of the SiCAL are displayed in Table 20.

| Tunsten-silicon sampling calorimeter | |
|---|---|
| Number of radiation lengths | 23.3 $X_0$ |
| Sampling (number of layers of 1.94 $X_0$) | 12 |
| Dimensions: <br>     Inner radius <br>     Outer radius <br>     Length <br>     z-position front face | 67.5 mm <br> 156.5 mm <br> 111.4 mm <br> 2503.6 mm |
| Weight of each detector | 105 kg |
| Detector active region: <br>     Inner radius <br>     Outer radius | <br> 69.5 mm (27.9 mrad) <br> 156.5 mm (62.7 mrad) |
| Detector crystals ($22.5^0$) <br>     per layer <br>     total | <br> 16 <br> 384 |
| Silicon crystal thickness | 300 μm |
| Total thickness of detector layer (G10 + crystal) | 1.8 mm |
| Thickness of aluminium support layer | 0.5 mm |
| Pad layout: <br>     Radial interval <br>     Azimuthal interval <br>     Successive layers rotated by (in ϕ) | <br> 16 x 5.3 mm pads <br> 2 x $11.25^0$ per crystal <br> $3.75^0$ |
| Readout | 2 x 6144 = 12288 channels |
| Energy resolution | 22.5 % / √E = 3.3 % @ 45 5 GeV |
| Phi resolution | $0.3^0$ ($R_{inner}$) to $0.2^0$ ($R_{outer}$) |
| Theta resolution | $0.085^0$ |
| Radial resolution | 0.4 mm |

Table 20: Parameters of the Luminosity Calorimeter, SiCAL

Beam energy measurements with an accuracy of (100-200) parts per million are needed for the determination of particle masses, including $m_t$ and $m_H$. At ILC Energy measurements both upstream and downstream of the collision point are foreseen by two different techniques to provide redundancy and reliability of the results. Detailed parameters such as size, position, coverage, have to be carefully tuned for particular FCC-ee machine conditions.

## 5. Conclusions

In the light of what has been learned from the characteristics and performances of the LEP and LHC detectors, amended by the more recent studies done for the ILC experiments, and taking into account the differences between a circular collider and a linear collider, an "ALEPH-like" design is proposed for the ASAHEL, where the limited-streamer tubes would be replaced by Micromegas chambers. These Micromegas detectors have been considered by ILD for the readout of the TPC and by both ILC experiments for the hadron calorimeters. Large size Micromegas detectors are developed for the upgrade of the ATLAS muon system. The possibility of using them for electromagnetic calorimetry is now explored by RD51. For ASAHEL the dimensions and granularities of the sub-detectors will be optimized to provide the required accuracy for the precise measurements targeted by the FCC-ee physics program. If the relevance of an "ALEPH-like" detector design was demonstrated, ASAHEL would optimally balance simplicity, expertise concentration, synergy with LHC and ILC experiments.